\documentclass[a4paper,11pt]{article}
\usepackage{graphics,epsf,epsfig,graphicx}
\begin{document}

\title{ Dynamical friction on cold fractal gas clouds
\\ applications to disc formation}

\author{B. Semelin
\footnote{ semelin@gravity.phys.waseda.ac.jp}
\\ Department of Physics, Waseda University \\
 \and  F. Combes \\Observatoire de Paris, DEMIRM
}

\maketitle

\abstract{ It is likely that cold molecular clouds form at high redshift,
before galaxies. Considering these cold clouds instead of hot homogeneous gas 
as the main baryonic component of proto-galactic halos may affect several 
issues during galaxy formation. 
In particular, the baryonic matter loses angular momentum through
dynamical friction on the dark matter halo. In numerical simulations using
hot gas for baryonic matter, this gives rise to the so-called angular momentum 
problem.
In this work, we study the dynamical friction exerted on
cold fractal gas by a collisionless background (dark matter) through 
high-resolution
numerical simulations. First, we find that, for values of the parameters 
relevant during galaxy formation, 
the friction does not alter the morphology of the fractal, which is mainly
driven by internal dynamics. Then, we show that the presence of substructures 
and inhomogeneities in a body has little effect on the global value of the 
friction. Parameters such as branching ratio and fractal
dimension also have little effect. In fact, we find out that the main effect 
comes from the
deformations and fluctuations of the fractal structure in cold gas clumps.
If the deformation time is of the same order of magnitude or shorter than the 
typical build-up time of the friction, the friction is weakened. We argue
that this effect is relevant for galaxy formation, and that the angular momentum
problem should not be worsened by including the small scale inhomogeneities
of the cold gas which occur at a resolution out of reach of present simulations.}


\section{Introduction} 

Our goal in this work is to investigate the consequences of 
including cold fractal gas in galaxy formation, with a focus 
on the angular momentum problem.

{ In the context of galaxy evolution, the importance of considering a multiphase 
medium for a realistic 
description of the interstellar medium has been first established by Cox \& 
Smith (1974) and McKee \& Ostriker (1977). They showed that the
temperature range relevant to galactic gas dynamics extend from 10 K up to
$10^6$ K. As for the density range, it extends from $10^{-2}$ cm$^{-3}$ in hot 
regions to possibly $10^{6}$ cm$^{-3}$ or more 
in the cores of molecular clouds (see, e.g.,
Faison et al. (1998) for detection of small dense structures).}
At each end of these ranges, hot plasma and cold
molecular clouds are driven by very different and complex physical processes,
and cannot be numerically described in a single phase.
For examples of simulations with multiphase descriptions of the interstellar
medium on full galactic 
scale see Rosen \& Bregman (1995), Gerritsen \& Icke (1997)
and Wada \& Norman (1999). These works focus on star formation and 
interstellar medium morphology, but not on galaxy formation.

Until now, most numerical studies of galaxy formation have 
considered
a single gas phase, treated through SPH dynamics, cooling being artificially
stopped at $10^4$ K (e.g., Weil et al. 1998, Thacker \& Couchman
2000 or Kay et al. 2000). 
{ Some authors have considered 
simple multiphase models (Yepes et al. 1997, Hultman \& Pharasyn 
1999) including cold gas. However, they chose to run simulations on
scales of several Mpc, which limits the resolution to $\sim$ 1 kpc. Important
differences in the physics of cold and hot gas appear below this scale. This is
the case for the fractal structure of cold gas clouds, which we focus on 
in this work.}

In simulations of galaxy formation, one of the
main difficulties arises from the "angular momentum problem". During
infall, the gas loses angular momentum through dynamical friction on the
surrounding dark matter halo: the resulting angular momentum of the disc
is too small by a factor of 10, compared to typical values for present days
galaxies (Navarro et al. 1995). 
Various solutions have been 
investigated to alleviate the problem. For example, Navarro \& Steinmetz 
(1997) 
used a photo-ionizing UV background to heat the gas in order to limit the 
collapse of the disc. They
found out however that this actually exacerbates the angular momentum problem,
since UV heating mostly prevents late infall of matter, which otherwise 
supplies most of the angular momentum. Along the same line,
Weil et al. (1998) showed that the angular momentum problem can be 
avoided if the gas is prevented from cooling at early epochs and until z=1.
However, they do not specify what process is responsible for suppressing the 
cooling.
Dominguez-Tenreiro et al. (1998) included simple star formation without 
feedback. They showed that the formation of a stellar bulge stabilizes the disc and 
reduces the infall of gas, thus limiting the loss of angular momentum. 
Thacker \& Couchman (2000) implement feedback from supernovae in 
galaxy
formation. They find that the final angular momentum can be increased through
feedback, but this effect may be insufficient to reach observed values. 
Consequently, the solution of the problem is still a matter of debate.
{ Moreover, the mass resolution is at best $10^6$ solar masses per particle
in usual galaxy formation simulations. This is not enough to take into account 
the fractal structure of cold gas and how dynamical friction acts on it.
A local study of this phenomenon is needed.}

Our approach is to consider the possibility of disc formation by the 
accretion of cold fractal gas. 
It is likely that clumps of molecular gas and GMCs
form very early, even before galaxies (Combes \& Pfenniger 1998).
The formation of such structures requires two conditions:
efficient early cooling leading to  quasi-isothermal collapse of
primordial density inhomogeneities; and inefficient star formation.
Indeed, the so called over-cooling problem is actually a problem only if the
cooling of the gas results in over production of stars at high redshift, which
is inconsistent with observations.
Evidence for both factors is presented in Combes \& Pfenniger
(1998) and references therein.
In the formation of discs from cold fractal gas, one essential process 
to study is the loss of angular momentum through dynamical friction.

Dynamical friction is a complex non-linear phenomenon, and it is a challenge to
predict its effect on a fractal medium. Chandrasekhar (1943) gives the first
evaluation of the friction on a test point-particle in a non self-gravitating
background. 
{ If $M$ is the mass of a test particle, $v_M$ its velocity, $m$
the mass of the background particles and $f(v_m)$ an isotropic distribution 
function for background particle velocities, the test particle velocity
changes according to

$$ {d \hbox{v}_M \over dt} = - 16 \pi^2 G^2 \ln \Lambda \,\,m (M+m)
\int_0^{v_M} \!\!\!\!f(v_m) v_m^2 dv_m  \,{\hbox{v}_M \over v_M^3} $$

\noindent
where

$$ \Lambda = {b_{max} V_0^2 \over G (M+m)}.$$

\noindent
Here $b_{max}$ is the maximum impact parameter (often the typical size of the 
background medium) and $V_0$ is the typical velocity difference 
between background and test
particle. There is an amount of uncertainty when assigning a value to
$\Lambda$ for a given physical application. In the case of an extended body,
Chandrasekhar's formula overestimates the friction since background particles with
small impact parameter are not as much deflected as in the case of a point 
test particle. For an extended body, it is
possible to define a minimum impact parameter $b_{min}$ and redefine the 
Coulomb parameter $\Lambda$:

$$ \Lambda={b_{max} \over b_{min}}$$

See Binney \& Tremaine (1987) for more details.
In the Chandrasekhar definition, dynamical friction is an acceleration. 
In this work, we will measure the dynamical friction on a given body as the 
net acceleration from gravitational forces 
exerted on this body by the background. We will refer to the norm of this
acceleration as the intensity of the friction.}

Binney \& Tremaine (1987) review various applications of
the dynamical friction. Important issues have been
to determine whether the effect is local or global, and whether the self-gravity
of the response is important or not.
Chandrasekhar's approach has a pure local formulation, which does
not explain the friction at a distance (Leeuwin \& Combes 1997).
This is essential in the braking of two interacting galaxies, for example.
In the case of a sinking sattelite, including the barycentric motion of the
primary body which acts as a background significantly reduces the dynamical 
friction (Weinberg, 1989).
Including also the self-gravitation of the primary body does not appear to 
produce large corrections (Bontekoe \& van Albada 1987; Zaritsky \& White 1988),
Another important point, rarely taken into account, is the
tidal deformation of the secondary component, the component
moving among the primary (background) particles. Since the friction
is due to the deformation of the interacting bodies, the freezing
of the smallest one leads to an underestimate of the friction; indeed it 
is the one which actually undergoes the largest deformation 
(cf Prugniel \& Combes 1992).
The Chandrasekhar formula can be used in many circumstances
for dimensional relevance, but its quantitative
value can be off by more than an order of magnitude, and this
uncertainty is hidden in the Coulomb parameter $\Lambda$
(e.g. Bontekoe \& van Albada 1987).
Most frequently, a collisionless background is assumed to
estimate the friction, although recently Ostriker (1999)
has estimated that the friction against a gaseous medium (collisional)
is much stronger. Her treatment is from linear perturbation theory
and does not take the self-gravity of the wake fully into account.

Our aim is to establish whether new parameters
affect the intensity of the friction when it acts on a fractal body.
We consider only a collisionless background, i.e. composed
of dark matter particles.
Our approach is through numerical simulations.

In section 2 we briefly present the code and numerical methods. In section
3, 4 and 5 we present the results from our simulations. 
We give our conclusions in section 6.

\section{ Numerical techniques and general setting}

\subsection{The code}

The code used to run the simulations is a parallel version of the tree-code 
(Barnes \& Hut 1986)
running on a Cray T3E at IDRIS. The present
version is a MPI-MPI2 adaptation of the original PVM-SHMEM version developed by
S. Ninin (1999). The code uses a binary tree technique and the
associated natural domain decomposition (Ninin 1999).
All simulations are run with an opening angle $\theta=0.8$,
governing the use of multipole expansion for a group of
particles, or their splitting in sub-groups . This version of the code uses a single 
time-step.
Our simulations require two types of particles: dissipative gas particles and 
collision-less dark matter particles. Dissipation, included in the simulations of
section 3 only, is implemented by inelastic
collisions, using the sticky-particle model 
(Levinson \& Roberts 1981). We use periodic boundary 
conditions throughout the paper, the potential being computed with Ewald method
(Ewald 1921). No cosmological expansion effect is included.

\subsection{ Geometry and scales: simulation setup}

{
Our goal is to reassess the angular momentum problem when the basic 
proto-galactic material is a distribution of cold fractal gas clouds. 
Studying such a system
presents a major difficulty from the computational point of view: the relevant 
scale range goes from $10^{-4}$ pc (the size of the smallest structures in
giant molecular clouds, Faison et al. 1998) to $10^5$ pc, a possible 
value for the 
size of the local proto-galactic halo. We have to reduce this range drastically 
in the simulations. The fractal structure of the cold gas mostly appears at 
scales smaller than 1 Kpc; at larger scales it is broken by the shear 
arising from the global geometry of the system (for example in differentially 
rotating discs).
Is dynamical friction on structures at this scale likely
to be a large part of the overall loss of angular momentum? It depends
on the merging history of the forming galaxy. Indeed, dynamical friction is
proportional to the mass of the body. So, if the galaxy formed through the
merging of a few large bodies, the friction at small scales will not play a 
large
part in the total angular momentum loss. However, if the merging history is
quiet, and the galaxy acquires matter mainly through accretion, then small-scale
friction must be evaluated as it may play an important part in angular momentum 
loss.   
Consequently, we focus on the scale of a single,
or a few molecular clouds; that is, a value of the order of 1 kpc for the
simulation box size.
Accordingly the baryonic mass in the box will be around $10^7$ solar
masses, and the dynamical time around 10 Myr. In the simulations these
quantities are normalized to 1.

The main effect of this zoom-in is that the loss of angular momentum
will actually be studied as a loss of linear momentum, as the difference in
in the bulk motion of the dark matter halo and the proto-galactic gas cloud
translates, on small scales, into a uniform drift between the two components.
Since we do not take the global dynamics into account, the drift 
between dark matter and gas must be artificially sustained. 
The drifting speed must be chosen of the same order as the virial velocity of 
the system ( $\sim 100$ km.s$^{-1}$ for a typical galaxy).

At the chosen scales 
(between 10 pc and 1 kpc), the gas has already
collapsed in dense clumps, raising the density ratio between gas and dark matter above one. 

}

\subsection{Periodicity and reference frame}
Since we set the simulation box size to sub-galactic scale, two questions arise:
how is the matter outside the simulation box taken into account, and what kind
of reference frame is associated with the simulation box, given the globally 
rotating nature of the proto-galactic cloud?

First, let us sum up our approach for the forming galaxy. We consider a
large number of fractal clumps of cold gas, accreting to form a disc, and we 
investigate the effect of
dynamical friction over scales smaller than  1 Kpc.
In this picture, the dynamical friction on each clump is local. 

The presence of matter (both gas and dark matter) outside the box is taken
into account by using periodic boundary conditions. This recreates the
distribution of clumps, albeit with an artificially regular pattern.

For a fully self-consistent simulation, a non-inertial frame should be used.
There is a well known procedure, for example, when simulating a small piece of 
a rotating disc ( Wisdom \& Tremaine 1998, Semelin \& Combes 2000).
However, we believe that in our study, Coriolis forces and drag from 
differential rotation are secondary processes since the dynamical friction 
arises from local gravitational effects. The  centrifugal force is of course
balanced by the global gravitational field of the forming galaxy. 
Consequently we use an inertial frame.

\begin{figure*}[t]
\centering
\includegraphics[width=17cm]{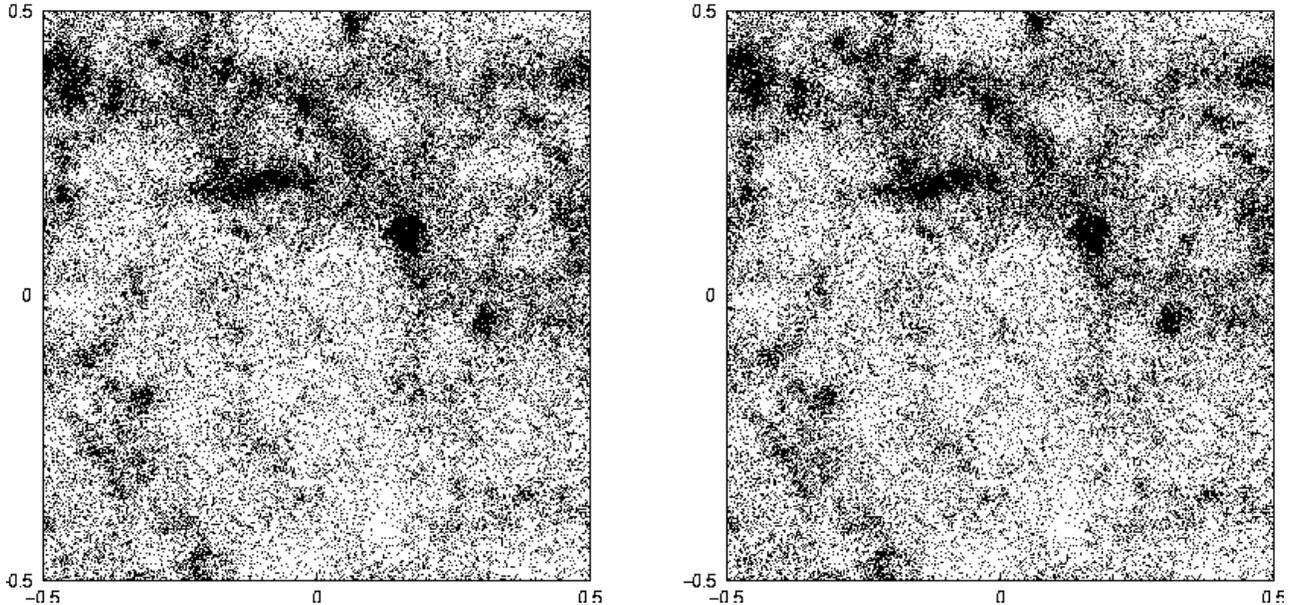}
\caption{Plots of the gas component (64000 particles) after 0.45 dynamical
times. On the left, the dark matter background (also 64000 particles, not plotted)
has no initial drift. On the right, the initial drift is $v_x=0.5$, $v_y=0.2$ and$v_z=0.2$, creating a dynamical friction. The density fields of the gas are remarkably similar.}
\label{dyn}
\end{figure*}

\section{Dynamical simulations}

\subsection{Initial conditions}

In this first set of simulations, we study the general case, a collapsing
gas medium within a drifting dark matter background.

The  average ratio of the density of the gas to the density of the dark matter
is set to 9. We use 64000 particles for each component. 

We want an homogeneous initial gas density field with fluctuations whose 
power spectrum follows a power law of index -2. { Using the continuity equation
in the linear regime, we can find the correct power law for the power spectrum
of the velocity field which produces the desired density fluctuations. We use
this field to assign velocities to the particles initially 
distributed on a regular grid.}
The velocity field amplitude is
set such that the simulation box includes several tens of Jeans masses for
the gas, allowing for efficient fragmentation. Furthermore, the gas undergoes 
dissipation in inelastic collisions. The dissipation is strong enough to 
absorb the potential energy released in the collapse (isothermal collapse 
condition). This means that at least half of the kinetic energy of the gas is
dissipated within 1 free fall time. 

The dark matter density field is homogeneous, and the thermal motion strong 
enough to prevent collapse on the box scale. We run two simulations. In the
second simulation, an additional uniform velocity
field is added to the dark matter to account for the drift between dark matter 
and gas responsible for the dynamical friction. Initially the
dark matter ,not the gas, is drifting since our frame is attached to the
gaseous component. Consequently we expect the dark matter
to slow down and the gas to accelerate until the drift is canceled as an
effect of dynamical friction.
The global drift velocity of the dark matter is chosen of the same order as the
ratio between box size over dynamical time (1 in the simulation dimensionless units, and of the order of $100$ km s$^{-1}$ in physical units).
This is also the order of magnitude of the virial speed for the dark matter 
component. 
According to Chandrasekhar's formula, it is the range where the friction is 
maximal (see also Ostriker 1999).

We run the two simulations over 1 dynamical time (typically 10 Myr)
using 400 time-steps, keeping energy conservation within 1 \%.

\subsection{Results}

Fig. 1  shows plots of the gas component after 0.45 dynamical time for 
both simulations. The plots are remarkably similar.  To quantify this, we
have binned the density field on a $16^3$ grid (each cell contains 15.6 gas
particles on average). Then, we define the usual mathematical norm:

$$ \| \rho \| \,=\, \left( \int \rho^2 \hbox{d {\bf x}} \right)^{1 \over 2}. $$

If we call $\rho_0$ the initial density field, $\rho_1$ the final density field
without dynamical friction, and $\rho_2$ the final density field with dynamical
friction, we have computed:

$$ {\| \rho_2-\rho_0 \| \over \| \rho_0 \|} = 1.90 \ldots$$

$$ {\| \rho_2-\rho_1 \| \over \| \rho_0 \|} = 0.091 \ldots $$

\noindent
For comparison we started from different initial conditions and 
called $\rho_3$ the collapsed state at t=0.45. We get:

$$ {\| \rho_3-\rho_1 \| \over \| \rho_0 \|} = 2.52 \ldots $$

\noindent
This shows that the two configurations in Fig. 1 are indeed similar. 
The effect of dynamical 
friction on the gas component is obviously
negligible at this stage. Later times in the integration do not show much more effect, 
while the fractal structure begins to be broken by the lack of energy input and
merges to a few "black holes". 
This does not mean that friction does not operate.

Fig. 2 shows a noticeable effect on the dark matter, which
loses a sizeable fraction of its initial momentum in the simulation including
a drift of the background over 0.8 t$_{dyn}$.
At $t=0.45 t_{dyn}$, however, the dark matter has lost only a few percent of its momentum. 
The dynamical friction acts, but not soon enough to affect the collapse of the 
gas. On the chosen scale, the dynamics of the gas cloud is 
dominated by its self-gravity which acts on much shorter time scales than the 
dynamical friction. Therefore, to study the effect of the dynamical friction, it 
is necessary to explore longer time scales in a quasi-stationary regime.

In this regard, the main difficulty is to obtain a long lived fractal structure
self-consistently. It requires full modeling of the interstellar medium
physics over a large range of scales, including complex processes for the 
dissipation and the energy input.
See, for example, Semelin \& Combes (2000) and 
Huber \& Pfenniger (2001). Work is still going on in this area. 
In this work, we will follow a simple approach. We freeze the fractal and 
run a semi-dynamical simulation. Since
the internal dynamics of the fractal are dominated by its self-gravity,
dynamical friction should not strongly affect the morphology of the 
fractal, and little physical significance is lost in a semi-dynamical 
simulation, as far as evaluating the intensity of the dynamical friction is 
concerned. In section 5 we
will follow another possible approach: imposing artificial dynamics on the
fractal.

\section{Dynamical friction on a frozen fractal body}

In this section we focus our attention on an individual structure in the
cold gas halo. This structure may be considered either as an individual giant 
molecular cloud or as a substructure embedded in a larger fractal structure. 
In both cases periodicity provides the neighboring gas structures. 
In physical systems, the slow decay of the orbit of the structure
during disc formation sustains the drift between gas and 
dark matter, and produces a local quasi-stationary state for the structure. 
Consequently, in the simulation we can attach the frame to the center of
gravity of the structure and get an almost inertial frame (centrifugal
force is balanced by gravity and Coriolis forces are negligible).
Since our simulation is local, the drift has to be artificially sustained.

\begin{figure}[t]
 \resizebox{\hsize}{!}{\includegraphics{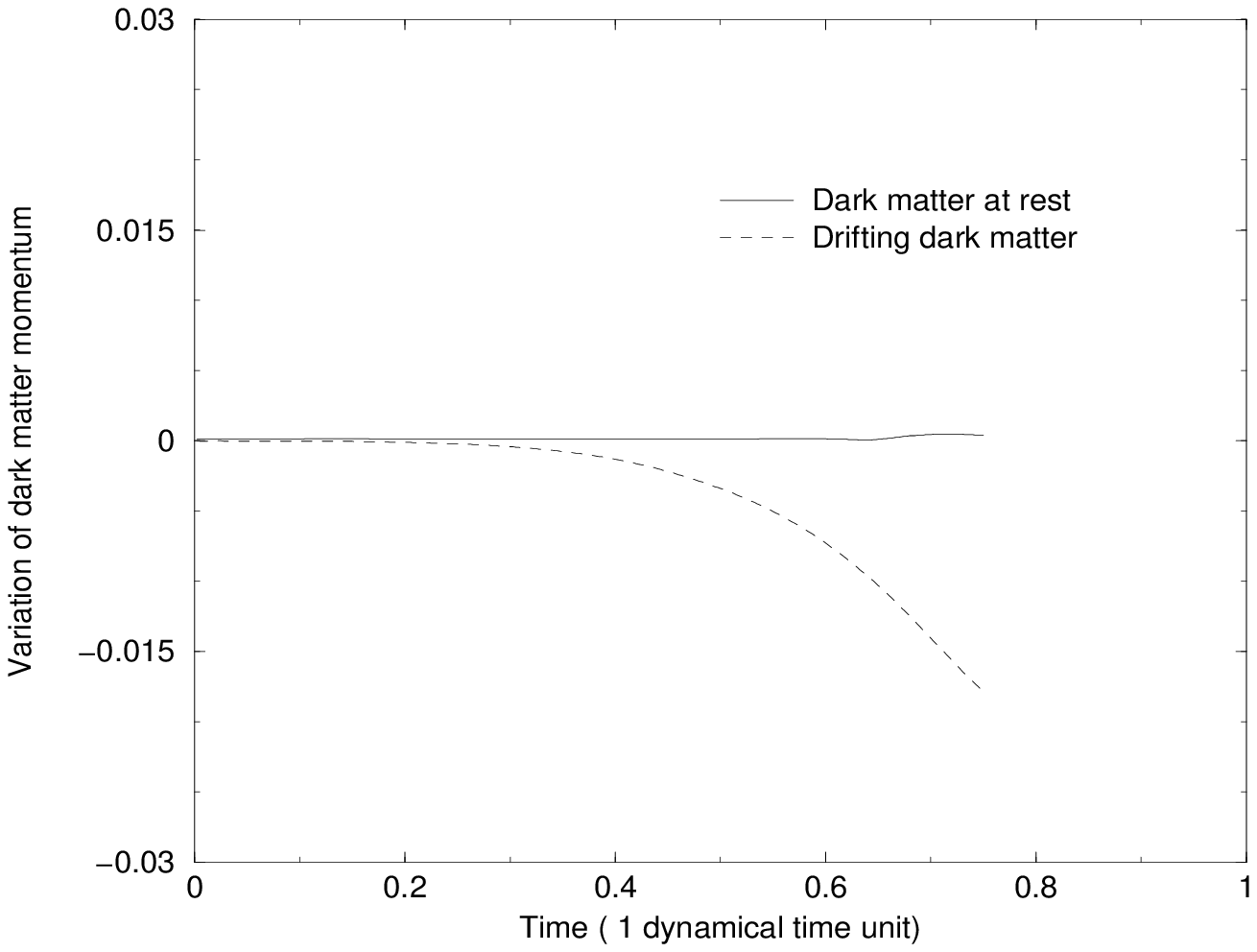}}
  \caption{Variation of the total momentum of dark matter as a function of time,
with or without dynamical friction (i.e. with or without a velocity
drift between gas and dark matter).
In the case where there is a velocity drift between the two components,
the dark matter total momentum shows a variation, giving
the order of magnitude of the friction. The fact that the variation is a decrease
is only due to the choice of reference frame of the gas, where the dark matter
is moving and the gas is at rest. In the galaxy's environment, the gas is
in motion relative to the dark matter, whose angular momentum increases.}
  \label{dmm}
\end{figure}

In the following simulations, a rigid, motionless multi-particle body of 
typical radius 0.1
(box size equal 1) and mass 0.2 (total mass equal 1) is subject to 
dynamical friction from a collisionless background. We give to this central
body an artificial fractal structure (described below).
This permits a meaningful comparison with
an equivalent homogeneous body and a fine control over the characteristic
parameters of the fractal structure.
Because of the periodic boundary conditions the maximal impact parameter is 
intentionnally much smaller
than in usual studies of the dynamical friction: around 3 to 4 times the
value of the minimal impact parameter depending on
exact definitions. This value is consistent with both possibilities for the
central body:
if it is a giant molecular cloud, neighboring clouds are likely to be
close by as observed in galactic discs, and if it is a substructure within a
fractal clump, typical substructure separation is also quite small.

The central body is much more dense (50 times) than the background dark matter, as can be expected for a dense 
molecular cloud. The total integration time of the simulations is about 40 
times the dynamical time the central body would have if it could evolve. 

\subsection{Building an artificial fractal body}

{

There are many possible procedures for building a fractal set of point particles.
A simple approach is to build a hierarchy of structures recursively. In this 
work we use a hierarchy of spheres. At level 0 of the hierarchy, the basic
sphere has radius $r_0$. Then we define two parameters. The branching ratio
$a$ specifies the number of sub-spheres we will include in each sphere. The
scale ratio $b$ defines the radii of the sub-spheres: at level $n$ spheres will
have a radius $r_0 b^{-n}$. To avoid a strong scale periodicity (see appendix A
in Semelin \& Combes 2000), we
introduce a variance when assigning the radius of each sub-sphere, equal to {$1 \over 3$} of the average radius at this level.
Given these two parameters, $a$ and $b$, the
recursive procedure is the following: each sphere is subdivided into $a$ 
sub-spheres
of radii $\sim b$ time smaller. We place the center of each sub-sphere at random
on the surface of the initial sphere. Different choices are possible here. For
example using a probability function  (${1 \over r^2}$, constant, etc) to 
distribute the centers of the sub-spheres inside the sphere. However, the
branching ratio being usually smaller than 10, the statistical realization of
the probability law is noisy, and different laws produce similar fractals.
When the center and radii of the spheres at the last level of the hierarchy have
been defined, we distribute particles in these spheres evenly.  
We divide each structure into substructures of equal mass. Here
again different choices can be made, for example using a Gaussian probability
function to determine the mass of each sub-sphere. Following this procedure, we 
build a fractal whose fractal dimension is defined by $a$ and 
$b$ according to: $D_f= {\ln(a) \over \ln(b)}$.
}

\begin{figure*}[t]
\resizebox{\hsize}{!}{\includegraphics{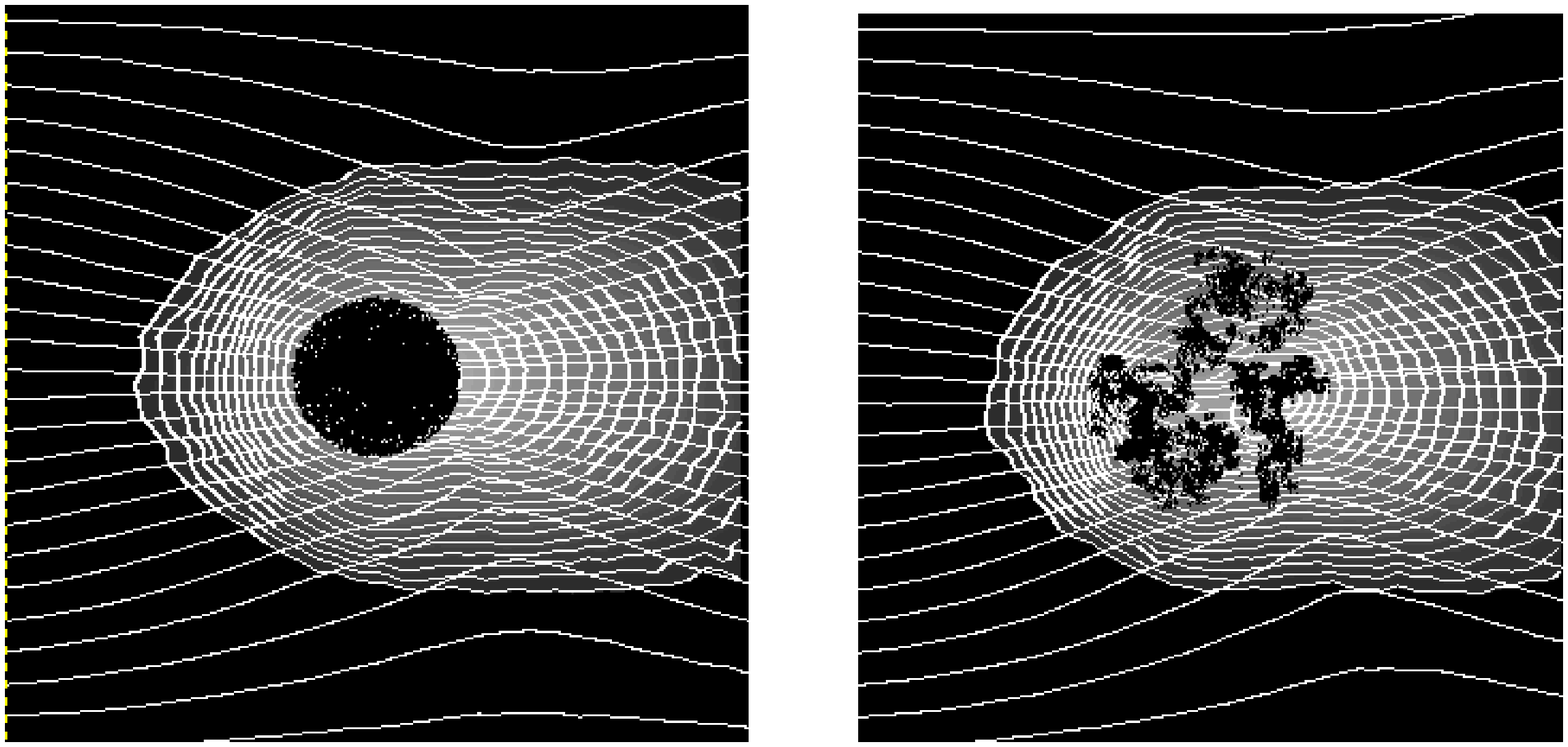}}
\caption{Stationary state dynamical friction on two different bodies. The
density contours and streamlines for a slice of the background are plotted,
smoothed over one crossing time. The body on the left in a homogeneous sphere.
The body on the right is a 5 level artificial fractal with dimension 2.}
\label{morph}
\end{figure*}

\subsection{Conditions of the simulation} 

The central body is typically made of 16000 particles. Its structure
ranges from a simple spherical distribution to an artificial fractal based 
on spheres with a 5 level hierarchy of structures, built using the procedure 
described in sec. 4.1 . Since the body is frozen, 
velocities are zero at all times.

The drifting dark matter background is made of collisionless particles. Two
sets of simulations have been run, the first with $30\,000$ background particles,
the second with 1 million background particles. Initially, particles are
distributed homogeneously with 
enough thermal motion to prevent collapse, and an additional drifting speed
in the $x$ direction with a value of $0.5$.
To reach a stationary
state, it is necessary to sustain the drifting speed of the background
against slowdown caused by dynamical friction. This is done in two steps.
First, particles exiting the simulation box at $x=0.5$ reenter at $x=-0.5$
with random $y$ and $z$ coordinates, and a regenerated momentum. However, simulations show
that this is not enough to reach stationarity. It is necessary to add the second
factor, an uniform force field which maintains the overall momentum of the
background. The norm of the force is proportionnal to the difference beetwen 
the actual background total momentum and the desired value.
This field accounts for the external large scale phenomena which
sustain the drift. Our simulations run over 2.5 box-crossing times
for the background.

\subsection{Morphology of the wake}

Fig. 3 shows simulations with a one million particle background, and
two different central bodies: a homogeneous sphere and a fractal of dimension 
$D=1.75$ (typical observed value for the interstellar medium, see 
Combes 1999), branching ratio $a=7$, and a scale ratio $b=3$. 
The density and velocity fields of the background are binned on a $32^3$ grid,
then averaged over one crossing time. 
Then we compute the density contours and the streamlines of the central slice
(x-y plane), and plot them on Fig. 3.
The complete central body is projected on the slice. 
In the case of the fractal body, the density contours of the background
show two local maxima, connected to substructures in the fractal. These denote
the existence of secondary gravitational wakes around the clumps within the
central body. Different slices (not plotted here), centered on different
substructures in the z direction show different maxima in the density contours
of the background, denoting more ``sub-wakes''.
These secondary wakes underline the fact that temporary structures form in the
dark matter around the density structure of the fractal body 
{\sl at all scales}.
These density structures are connected with perturbations in the velocity
field, stirring velocity dispersion at all scales. Confirmation of this
effect is shown in Fig. 4, which plots the average difference of
the friction intensity between two particles in the central body as a function
of their distance. This diagnosis is plotted for both the spherical body and a 
5 level fractal. Obviously, in the case of the fractal, the spatial 
fluctuations of
the friction remain strong at small scales, whereas they decrease in the case
of the sphere. This is again due to the presence of density structures producing
secondary wakes at all scales in the case of a fractal body. 
It is however obvious in 
Fig. 3 that the large-scale shape of the wake is very similar for 
the two kinds of bodies.

\begin{figure}[t]
\resizebox{\hsize}{!}{\includegraphics{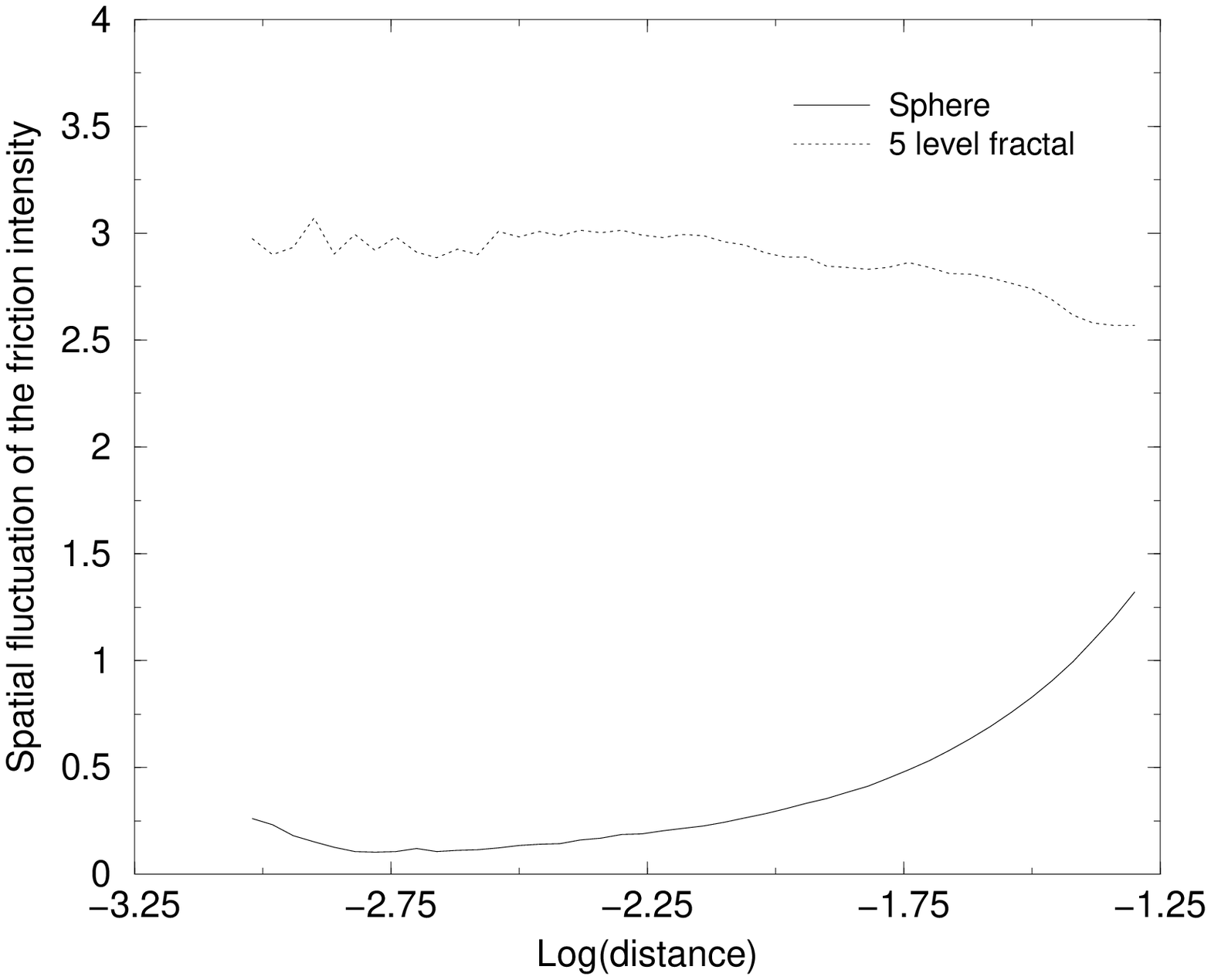}}
\caption{Fluctuation of the friction between two particles of the central
body, as a function of the inter-particle distance. We compute the difference
of gravitational force exerted by the background on two particles for every
pair of particles in the central body.
Then we average the norm of the difference over pairs with the same separation.
Plots are given for an homogeneous sphere and a 5 level fractal (a=7, b=3).
There are big
fluctuations in the case of the fractal. These average out at large scales,
where the friction intensity tends to that of an homogeneous sphere.}
\label{corel}
\end{figure}

\subsection{Total intensity of the dynamical friction}

We now assess the influence of the various parameters of the fractal body on the total 
intensity of dynamical friction, since it could affect the loss of angular
momentum in galaxy formation.

The most significant parameter is the scale range of structures present
in the fractal.
Does the presence of substructures affects the global intensity of the 
friction ?
In the case of our artificially constructed fractal, the scale range 
of the structures is controlled by
the number of levels in the hierarchy, from 1, the homogeneous sphere, to 5
levels for a fractal extending over two orders of magnitude in scale. 

\begin{figure}
\resizebox{\hsize}{!}{\includegraphics{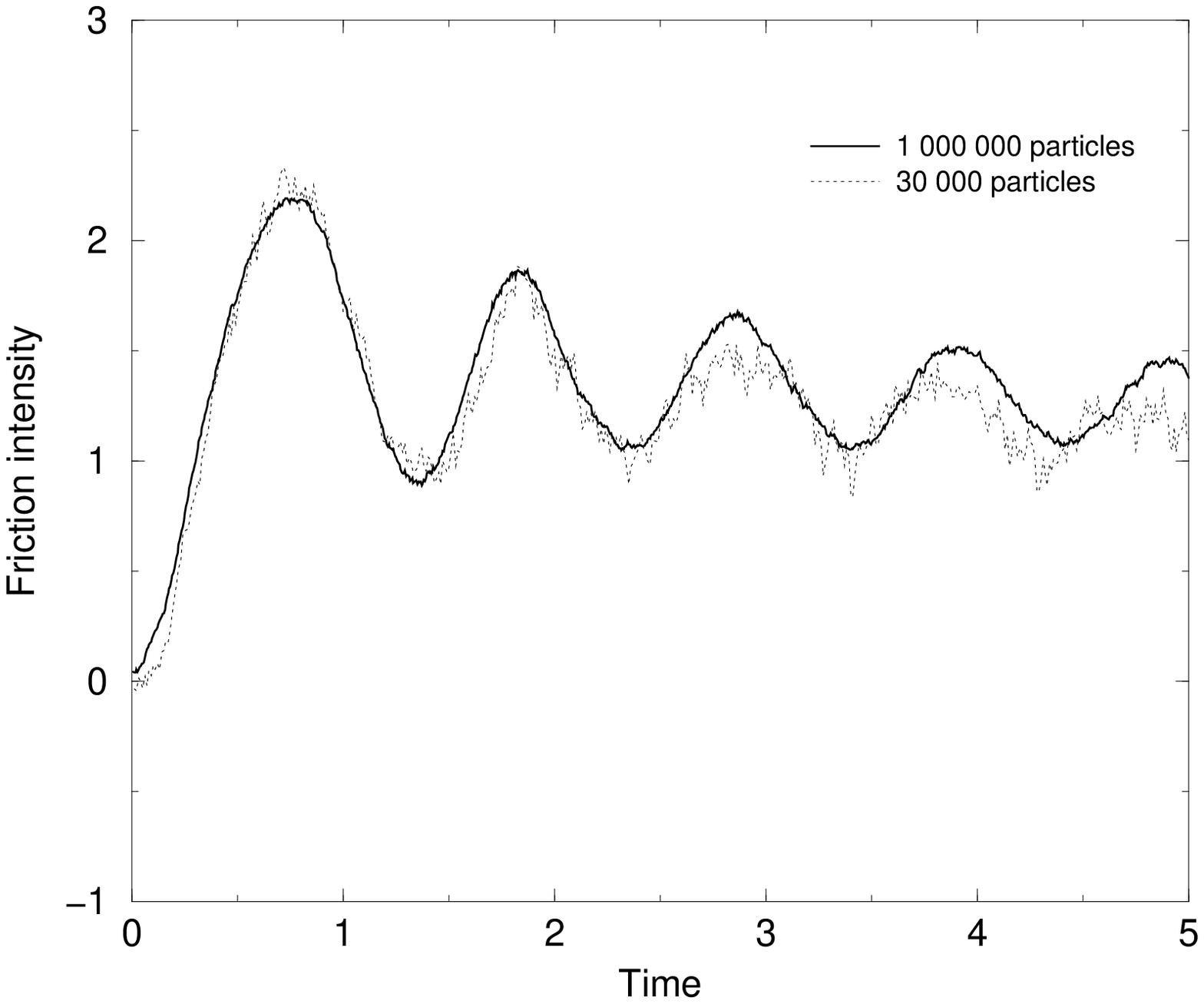}}
\caption{Total intensity of the dynamical friction for backgrounds with 30000
and 1000000 particles. The  curves are not smoothed. The central body is a
2 level "fractal"; that is, a composite object made of 7 spheres of varying 
radii.}
\label{rough}
\end{figure}

Fig. \ref{rough} shows plots of the intensity of the dynamical friction
as a function of time. The dynamical friction is evaluated by summing the
accelerations of the particles in the central body. The self-gravity of
the body cancels out. The two curves correspond
to a 1 million and a 30 000 particle background. An important noise reduction
can be observed, calling for the use of 1 million particle simulations. 
As can be expected, the observed damped oscillations have a period similar to 
the free fall time of the background.
We can see that stationarity  is not fully reached after 2.5 crossing times.
However, because of the computation cost of the 1 million particles simulations,
we will be satisfied to evaluate an average value for the friction by 
smoothing the later parts of the curves.

\begin{figure}
\resizebox{\hsize}{!}{\includegraphics{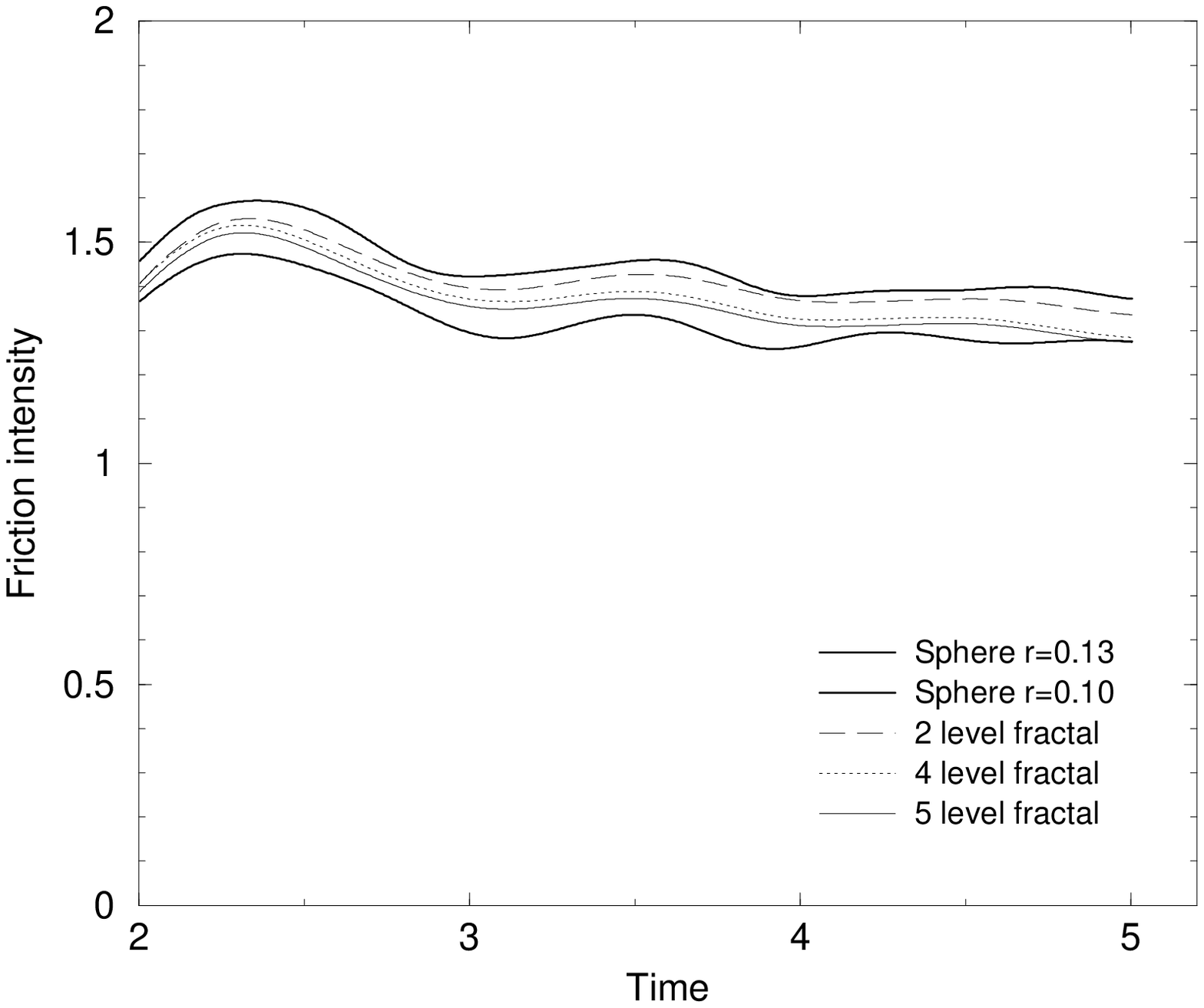}}
\caption{Total intensity of the friction as a function of time, smoothed over
one crossing time. Plots for 2 different spheres, and 3 different fractals (a=7,
b=3).}
\label{intensity}
\end{figure}

Fig.
\ref{intensity} presents smoothed plots of the intensity of the friction as a 
function
of time for 5 different simulations. All simulations use a 1 million particle
background, and extend over 2.5 crossing times with 1000 time steps per
simulation. The two
thick full lines correspond to spherical central bodies with radii 0.1
and 0.13, bracketing the values of the effective radii of the other bodies.
Indeed, when applying the Chandrasekhar formula to an extended body, the
effective radius determines the minimal impact parameter, and thus the intensity
of the friction. We must keep track of the variation due to this parameter.
The thin lines correspond to fractal bodies with 2, 4 and 5 levels in the
hierarchy. The obvious conclusion is that the presence of substructures has 
little effect on the total intensity of the friction.
Moreover, we have been able to verify that what little dispersion in the
intensity of the friction can be seen between the 2, 4 and 5 level fractal
has a simple orgin, for the most part. It can be partly accounted for using
Chandrasekhar's formulav by a 
variation of the total effective radius, i.e. a variation in the minimal 
impact parameter.
Let us emphasize again that the ratio of maximal to minimal impact parameter is
small, so inhomogeneities in the central body can be felt throughout the
simulation box, and this absence of measurable effect was not expected.

Complementary simulations have been run with 30 000 particle backgrounds to
evaluate the influence of parameters such as the branching rate in the fractal
and the value of the fractal dimension. As can be expected from the previous
result, no systematic effect has been detected other than from a variation
of the minimal impact parameter.

\section{Dynamical friction on a fractal with an artificial dynamics}

\begin{figure*}[t]
\resizebox{\hsize}{!}{\includegraphics{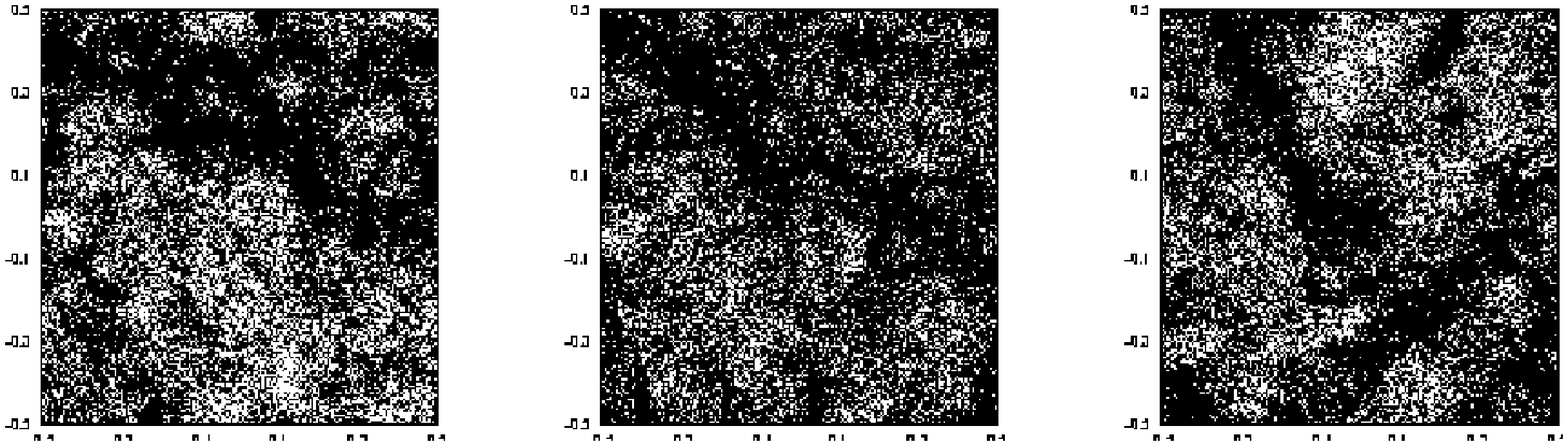}}
\caption{ Plots of gas configurations at different times in the artificial
motion. From left to right, the gas follows a linear transformation between
two fractal states resulting from self-consistent simulations.}
\label{film}
\end{figure*}

In the previous section we established that, beyond its size, the specific 
spatial structure of a body submitted to dynamical friction has little effect
on the intensity of the friction. However, in addition to the fractal
structure, cold molecular gas has another important property: the 
structures evolve and change shape constantly, as a result of dissipation and 
energy input processes. We will now investigate the impact of this behavior on
the dynamical friction.

\subsection{Conditions of the simulations}

As already stated, modeling the long-lived dynamical fractal 
structure of molecular clouds self-consistently
is still a challenge. Since our focus is on dynamical friction, we choose to
impose artificial dynamics on the fractal structure. Using the setup described
in section 3, we run two simulations, with two different random realizations of
the initial conditions. We record the fractal structure of the gas component at
$t=0.5 \,t_{dyn}$. These two self-consistently built fractal configurations 
are the basis of our artificial motion. 

One of the configurations is used for the initial positions of the gas 
particles. Then, each particle moves in a straight line with an individual 
constant speed toward its position in the second fractal. All particles reach
the second fractal configuration at the same time. The process is then
repeated toward a globally translated version of the first fractal, and so on
if necessary. Fig. 7 shows the two fractal states and one intermediate
state. Although the intermediate state is less clumpy, it retains 
some definite structures. The fact that the intermediate states do not come
closer to homogeneity is due to the relation between the positions of a 
given particle in the two extremal fractal states. Indeed, the simulations 
producing these states start from particles on a grid with a specified velocity 
field. In most cases closeness on the grid implies closeness of the positions
in the final fractals. This closeness is retained in part in the transition
between the two configurations.

The characteristics  of the background are the same as in section 3.

\subsection{Results}

We want to evaluate the effect of a shifting fractal structure on the friction
exerted by the background.

First, we run two reference simulations with two different frozen structures: 
one with the initial, fully fractal configuration and the other with the 
intermediate, 
less-clumpy state. Then we run two simulations with an evolving fractal 
structure, following the procedure described in sec. 5.1 for the gas motion. 
We choose two different values, 
0.5 and 1, for the time required to shift from one fully fractal structure to the other.
The dynamics of the background is self-consistently computed under the 
gravitational potential of the gas.

Fig. 8 shows plots
of the total dynamical friction exerted on the gas by the background for
the four simulations. As can be expected, the friction on the frozen less-clumpy
intermediate state is weaker than on the frozen, clumpy, initial state.
In both cases,
the friction builds up until $t=0.5$, then decays because of the 
heating-up of the wake. By comparison, the friction is weaker
for both simulations involving a shifting fractal. Indeed the
friction does not have time to build up around clumps as they move, dissolve and
collapse constantly. As the curves show, the faster the evolution, the weaker 
the friction. In the limit of very slow evolution, the friction intensity 
would reach the same value as for the frozen states.

In actual situations, this effect is relevant if the typical formation time of
the wakes, which is of the order of ${b_{\hbox{ \scriptsize max}} \over v_{\hbox{ \scriptsize drift}}}$
, is of the same order or
longer than the typical evolution time of the structure 
( $b_{\hbox{ \scriptsize max}}$ is the maximal impact parameter and
$v_{\hbox{ \scriptsize drift}}$ is the
drifting speed between the background and the body).
If we consider 
an average GMC of size 100 pc and speed dispersion 10 km s$^{-1}$, the typical
deformation time is 10 My. In the case of either galaxy mergers or accretion
of cold fractal clumps during disc formation we can evaluate 
$v_{\hbox{\small drift}}=
100 \,\,\hbox{km.s}^{-1}$ and $b_{\hbox{\small max}}$ is a few times the size of the clumps,
for example 500 pc. This gives a typical build-up time of the gravitational 
wakes of 5 My. For these values, the weakening of the dynamical friction
described above is relevant. The most direct application is to the angular
momentum problem in galaxy formation, which will not be worsened by the
presence of fluctuating small scale structures  in the cold fractal gas. 

\begin{figure}
\resizebox{\hsize}{!}{\includegraphics{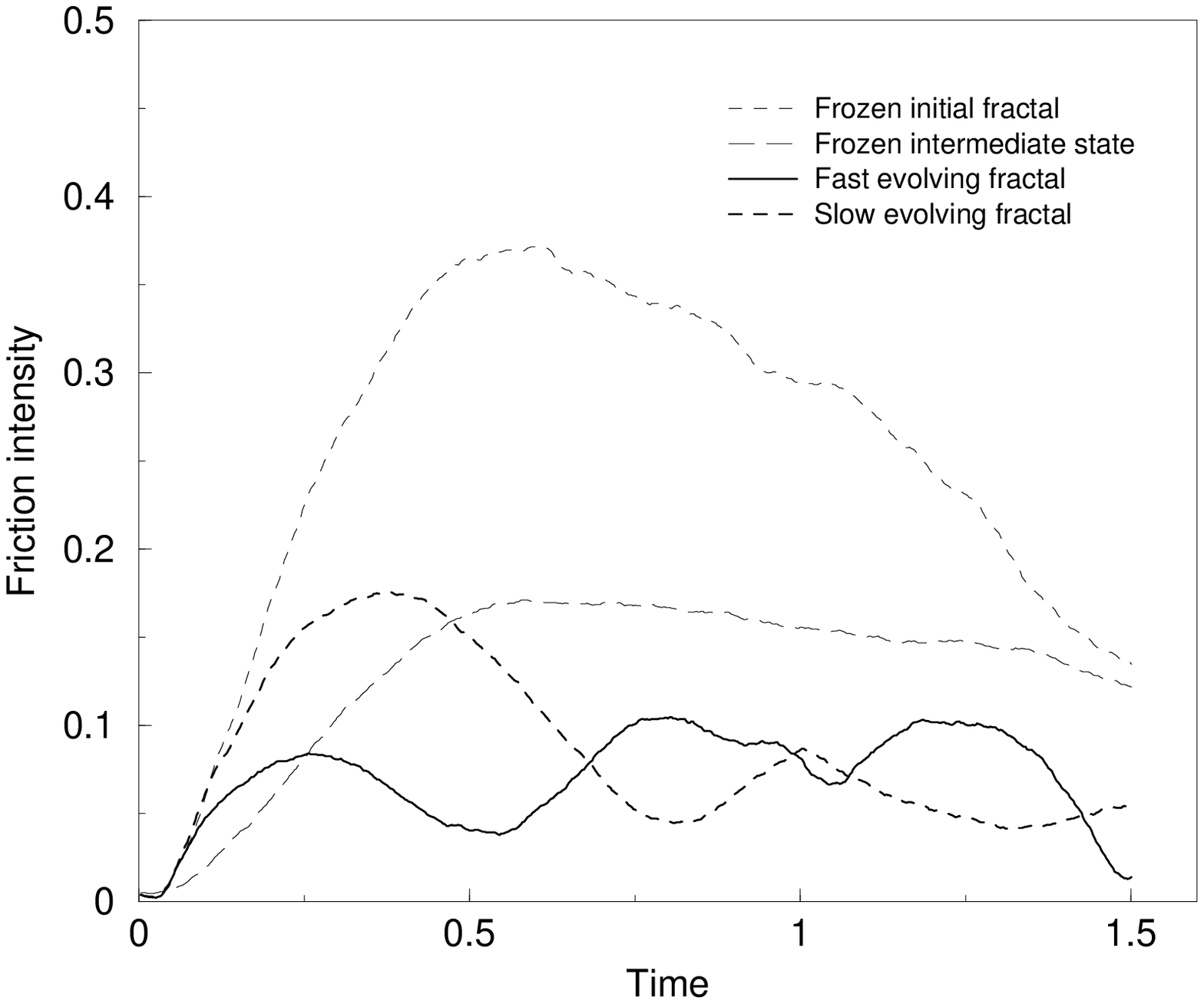}}
\caption{Time evolution of the dynamical friction in four different cases. Thin
lines are for friction on frozen objects, the initial fractal and the
intermediate state. Thick lines are for objects evolving from the first fractal,through the intermediate state, to the second fractal, and so on. The duration
of the transformation is 0.5 in one case, and 1. in the other.}
\label{fluct}
\end{figure}

\section{ Conclusion }

In this work, we have investigated dynamical friction
acting on fractal structures such as can be found in molecular clouds. We have
argued that this process can be relevant to galaxy formation, and more
specifically to the angular momentum problem. Indeed GMCs are likely to form
at high $z$ and play an important part in the formation of galactic discs, during
which angular momentum is lost through dynamical friction between the gas
and the dark matter halo.
{ Due to their different physics, the dissipative gas and the collisionless 
dark matter particles have different dynamics, and relative bulk motions
between these two components are expected to be of the order of
the virial velocities in galactic systems. These
relative motions should apply to galaxy 
formation conditions, gravitational collapse or galaxy mergers as well. }

We have first established that, for relevant values of the parameters, the
fractal structure of the gas is driven by internal processes including 
self-gravity, and that dynamical friction is not likely to be strong enough
to affect GMC morphology. Then we examined if and how GMC morphology 
could affect the intensity of the friction. Using a frozen central structure,
we have 
established that the presence of inhomogeneities and substructures has little 
effect on the intensity of the friction. This holds down to small values of
the maximum impact parameter, for which inhomogeneities are most likely
to affect the global shape of the gravitational wake. What we did find, however,
is the presence of secondary wakes connected to substructures. This shows
that friction on a fractal body stirs velocity dispersion on all scales in the
drifting background.

To complete the study, we have evaluated the response of the friction to
changes in the shape of the
fractal structure with time, as happens in molecular
clouds. We have shown that if the typical formation time of the gravitational
wake is longer or of the same order as the deformation time of the structures,
the friction is weakened. This effect can have important consequences for the
dynamics of galactic discs. If we consider the formation of galactic discs from 
an assembly of fractal clumps, the weakened friction on the dark matter
means a smaller loss of angular momentum for the gaseous disc. 

In numerical simulations of galaxy mergers or galaxy formation, particles 
representing gas clouds have very large mass, due to the lack of 
resolution and the small number of particles. Typically, a gas cloud is of 
the order of a GMC mass or bigger (Barnes \& Hernquist 1992).
The same is true for
particles representing stars, that are as massive as globular clusters.
Since the dark matter particles are of comparable mass, friction
at small scale cannot be computed and is ignored. However, the
friction is important at all scales; and the results presented here show
that the fine structure of the interstellar medium 
does not lead to a worsening of the problem.

{\small
Most computations in this work have been realized on the Cray T3E of 
the CNRS computing center, at IDRIS. The work of B. Semelin was supported by
a JSPS grant.
}

\end{document}